\documentclass[conference]{IEEEtran}
\IEEEoverridecommandlockouts
\usepackage{cite}
\usepackage{amsmath,amssymb,amsfonts}
\usepackage{algorithmic}
\usepackage{graphicx}
\usepackage{textcomp}
\usepackage{xcolor}
\usepackage{subcaption}
\usepackage{multirow}
\def\BibTeX{{\rm B\kern-.05em{\sc i\kern-.025em b}\kern-.08em
		T\kern-.1667em\lower.7ex\hbox{E}\kern-.125emX}}
\begin{document}
	\title{Mitigating Automotive Radar Interference using Onboard Intelligent Reflective Surface\vspace{-5mm}}
	\author{\IEEEauthorblockN{Shree Prasad Maruthi$^{*}$, Karrthik G. K.$^{**}$, Vijaya Krishna A.$^{\dagger}$, Mahbub Hassan$^{\ddagger}$, Jinhong Yuan$^{\ddagger\ddagger}$}
		\IEEEauthorblockA{$^{*}$Dept. of Electrical and Electronics Engineering, WILP Division, BITS Pilani, India \\
			$^{**}$Charcoal Technologies, Bengaluru, India, $^{\dagger}$Bosch Global Software Ltd., Bengaluru, India \\
			$^{\ddagger}$ School of Computer Science and Engineering, University of New South Wales, Sydney, Australia\\
			$^{\ddagger\ddagger}$ School of Electrical Engineering and Telecommunications, University of New South Wales, Sydney, Australia\\
			Email: shreeprasadm@gmail.com, karrthik.gk23@gmail.com, A.VijayaKrishna@in.bosch.com, mahbub.hassan@unsw.edu.au,\\ j.yuan@unsw.edu.au}
		\vspace{-12mm}}
	\maketitle
	\begin{abstract}
		The use of automotive radars is gaining popularity as a means to enhance a vehicle's sensing capabilities. However, these radars can suffer from interference caused by transmissions from other radars mounted on nearby vehicles. To address this issue, we investigate the use of an onboard intelligent reflective surface (IRS) to artificially increase a vehicle's effective radar cross section (RCS), or its ``electromagnetic visibility." Our proposed method utilizes the IRS's ability to form a coherent reflection of the incident radar waveform back towards the source radar, thereby improving radar performance under interference. We evaluated both passive and active IRS options. Passive IRS, which does not support reflection amplification, was found to be counter-productive and actually decreased the vehicle's effective RCS instead of enhancing it. In contrast, active IRS, which can amplify the reflection power of individual elements, effectively combats all types of automotive radar interference when the reflective elements are configured with a 15-35 dB reflection gain.
	\end{abstract}
	\begin{IEEEkeywords}
		Intelligent reflective surface (IRS), active IRS, millimeter wave automotive radar, frequency modulated continuous wave (FMCW), radar interference, advanced driver assistance system (ADAS).
	\end{IEEEkeywords}
	\vspace{-2mm}
	\section{Introduction}
	Automotive radars are quickly becoming a standard feature in modern cars due to their ability to operate effectively under inclement weather conditions such as fog, snow, and darkness~\cite{rad_adv}. This advantage over LIDARs and cameras has made automotive radar essential for driverless cars and the preferred choice for a range of advanced driver assistance system (ADAS) applications, including active cruise control, blind spot detection, cross-traffic alerts, and parking assist, among others, for existing cars~\cite{sae}.
	
	At present, frequency modulated continuous wave (FMCW) based automotive radar operating in the millimeter wave band is the preferred choice by original equipment manufacturers (OEM) due to its simple hardware implementation~\cite{fmcw_pop}. However, with the lack of standardization and the anticipated increase in the use of radars in future vehicles, the issue of interference has become a significant concern, as it can increase the noise floor and severely limit radar's target detection performance~\cite{rad_adv,fmcw_pop}. Numerous techniques for interference mitigation have been proposed, but they often require computationally demanding signal processing tasks, such as digital beamforming, adaptive nulling, and space-time processing~\cite{mosarim}.

	Intelligent reflective surfaces (IRS)~\cite{simris} have emerged as a new paradigm in wireless communication, with the potential to achieve significant performance gains. Their ability to reflect electromagnetic signals in the desired direction helps to overcome typical wireless channel limitations. Active IRS~\cite{activeris}, which features reflecting elements that provide amplification (reflection gain), can offer even further performance improvements.
	
	This paper explores a novel approach that utilizes an active IRS array to enhance the target detection performance of automotive radars. Specifically, we propose embedding an active IRS at the rear end of a target vehicle to significantly improve the strength of the backscattered signal in the direction of the radar. Our approach seeks to address the challenge of interference and limitations of radar systems in detecting targets. This paper makes the following contributions:
	

	$\bullet$ We propose the use of a rear-mounted onboard intelligent reflective surface (IRS) to artificially increase the effective radar cross-section (RCS) of a vehicle, enabling the detection of target vehicles under interference from nearby radars. To the best of our knowledge, this is the first proposal to combat automotive radar interference using onboard IRS.
	
	
	$\bullet$ We demonstrate that \textbf{passive} IRS, which does not support reflection amplification, is too weak to combat typical radar interference. In fact, our analysis reveals that covering the rear-end of vehicles with passive IRS would be counter-productive and could \textbf{reduce} the vehicle's effective RCS.
	
	$\bullet$ Our simulations show that an active IRS with 15-35 dB reflection gain can help achieve a probability of detection greater than 99\%. Therefore, even the most basic single-antenna radars can overcome various types of automotive radar interference without requiring complex signal processing tasks.
	
	
	The rest of the paper is structured as follows. Related work is presented in Section II. In Section III, the FMCW signal model is described. Section IV describes different interference scenarios. The automotive interference model is presented in Section V, followed by the IRS-assisted target detection in Section VI. Simulation results are presented and discussed in Section VII before concluding the paper in Section VIII. 
	\section{Related Work}
	Researchers have begun developing techniques to mitigate interference in automotive radars. Existing work in this area can be categorized as signal-processing-based, deep-learning-based, or IRS-based.
	
	\textbf{Signal processing and deep learning:} Due to the lack of standardization and increasing usage of FMCW-based automotive radars for various ADAS applications, the MOSARIM project~\cite{mosarim} was initiated by the European Commission to study different interference scenarios and suggest countermeasures. Several techniques have been proposed in the literature to combat interference, including repurposing the radar for both communication and sensing~\cite{radcom}, conventional approaches based on minimizing interference in the time and frequency domain~\cite{rad_lit1,rad_lit2}, and adaptive filtering techniques to estimate and cancel the interferer samples~\cite{rad_lit3}. Compressed sensing blindly mitigates interference~\cite{rad_lit4}, while supervised and unsupervised deep neural network techniques have been proposed~\cite{rad_lit5,rad_lit6}. However, deep neural network techniques require a large dataset under dynamically varying conditions and may not always provide robust parameter estimates under different types of interference~\cite{rad_lit7}.

	
	
	\textbf{Passive IRS:} The aforementioned approaches suppress interference and thereby improve the effective SIR. In contrast, an intelligent reflective surface (IRS) improves the SIR by coherently reflecting the signal from the target, thus enhancing its ``electromagnetic visibility" or effective RCS. Passive IRS has been investigated for improving the reliability of wireless communication in \cite{rs1,rs2,rs2_2}, while passive IRS-assisted radar, where the IRS is embedded in infrastructure, is studied in \cite{rs3} to extend sensing range and suppress interference. In \cite{rad_lit8}, passive IRS embedded on vulnerable road users is shown to improve their visibility when located at short range.
	
	\textbf{Active IRS:} Recently, active IRS, which amplifies reflected signals, has gained attention to further improve performance gains in wireless communication \cite{activeris}. Although active IRS is still in its infancy, it is considered as one of the key technology enablers for mmWave and terahertz communications \cite{rs5}. 
	In \cite{rs4}, an active IRS-based approach is proposed to improve radar detection capability under severe path loss. However, the IRS is not embedded on the target. In contrast, the active IRS proposed for automotive radar in this paper is mounted on the target vehicle and is shown to significantly improve the SIR under the challenging scenario of short-range interferer and long-range target.
	\section{FMCW Signal Model}
	In this section, we present the transmitted and received signals models for the FMCW radar. Notations used throughout this section are summarised in Table \ref{tab:tabel1p}.
	\subsection{Transmitted FMCW Radar Signal Model}
	The transmitted waveform by an FMCW automotive radar is a sequence of chirps of the form \cite{rad_dm}
	\begin{equation}
		x_{T}\left(t\right) = \sqrt{P_{t}} \sum\limits_{l=0}^{N_{f}-1}s_{b}\left(t-lT_{r}\right)\cdot a\left(t\right)
	\end{equation}
	where $s_{b}\left(t\right) = \exp\left[j2\pi\left(f_{c}t+\frac{1}{2}\frac{B_{s}}{T_{r}}t^{2}\right)\right]$ is a chirp signal, $a\left(t\right) = \text{rect}\left(\frac{t-T{r}/2 -lT_{r}}{T_{r}}\right)$, $P_{t}$ is the transmitted power, $T_{r}$ is the chirp duration, $f_{c}$ is center frequency, $B_{s}$ is the sweeping bandwidth, and $N_{f}$ denotes the number of chirps in a frame. Further, the slope $S_{V}$ of the chirp signal is defined as $\frac{B_{s}}{T_{r}}$. The target parameters and their resolutions measured by the radar with receiver bandwidth $B_{c}$ are given as i) Maximum range $R_{max}$ = $\frac{B_{c}c_{0}}{2S_V}$, ii) Range resolution $\Delta R$ = $\frac{c_{0}}{2B_{s}}$, iii) Maximum velocity $\nu_{max}$ = $\frac{\lambda}{4T_{r}}$, and iv) Velocity resolution $\Delta \nu$ = $\frac{\lambda}{2N_{f}T_{r}}$. Here, $c_{0}$ and $\lambda$ respectively denote the speed of light and wavelength.
	\begin{table}[h]
		\caption{Parameters and Notations}
		\resizebox{\columnwidth}{0.4\columnwidth}{\begin{tabular}{|cccc|}
				\hline
				\multicolumn{4}{|c|}{\textbf{FMCW and IRS Parameters}}                              \\ \hline
				\multicolumn{1}{|c|}{Symbol}               & \multicolumn{1}{c|}{Quantity}                                           & \multicolumn{1}{c|}{Symbol}                & Quantity                                                                  \\ \hline
				\multicolumn{1}{|c|}{$P_{t}$}              & \multicolumn{1}{c|}{Transmit power}                                     & \multicolumn{1}{c|}{$T_{r}$}               & Chirp duration                                                            \\ \hline
				\multicolumn{1}{|c|}{$N_{f}$}              & \multicolumn{1}{c|}{Number of chirps per frame}                         & \multicolumn{1}{c|}{$f_{c}$ and $\lambda$} & Center Frequency and Wavelength                                           \\ \hline
				\multicolumn{1}{|c|}{$B_{s}$}              & \multicolumn{1}{c|}{Sweeping bandwidth}                                 & \multicolumn{1}{c|}{$S_{V}$}               & Slope of victim radar.                                                    \\ \hline
				\multicolumn{1}{|c|}{$R_{\text{max}}$}     & \multicolumn{1}{c|}{Maximum measurable range of radar}                  & \multicolumn{1}{c|}{$\Delta R$}            & Range resolution                                                          \\ \hline
				\multicolumn{1}{|c|}{$\nu_{max}$}          & \multicolumn{1}{c|}{Maximum measurable velocity of radar}               & \multicolumn{1}{c|}{$\Delta v$}            & Velocity resolution                                                       \\ \hline
				\multicolumn{1}{|c|}{$c_{0}$}              & \multicolumn{1}{c|}{Velocity of EM wave}                                & \multicolumn{1}{c|}{$R$}                   & Range of target                                                           \\ \hline
				\multicolumn{1}{|c|}{$\nu$}                & \multicolumn{1}{c|}{Relative velocity of ego and target vehicle}        & \multicolumn{1}{c|}{$P_{r}$}               & Received signal power                                                     \\ \hline
				\multicolumn{1}{|c|}{$\tau_{l}$}           & \multicolumn{1}{c|}{Round-trip-time of $l^{text}$ chirp signal}         & \multicolumn{1}{c|}{$G_{t}$ and $G_{r}$}   & Gain of transmitter and receiver antenna                                  \\ \hline
				\multicolumn{1}{|c|}{$\sigma$}             & \multicolumn{1}{c|}{RCS of target}                                      & \multicolumn{1}{c|}{$B_{c}$}               & Receiver bandwidth/LPF cut-off frequency                                  \\ \hline
				\multicolumn{1}{|c|}{$f_{b}$}              & \multicolumn{1}{c|}{Beat frequency}                                     & \multicolumn{1}{c|}{$f_{D}$}               & Doppler frequency                                                         \\ \hline
				\multicolumn{1}{|c|}{$T_{u}$}              & \multicolumn{1}{c|}{Vulnerable time period}                             & \multicolumn{1}{c|}{$S_{I}$}               & Slope of interferer radar.                                                \\ \hline
				\multicolumn{1}{|c|}{$\tau_{I}$}           & \multicolumn{1}{c|}{Time offset.}                                       & \multicolumn{1}{c|}{$P_{t_{I}}$}           & Transmit power of interferer radar                                        \\ \hline
				\multicolumn{1}{|c|}{$G_{t_{I}}$}          & \multicolumn{1}{c|}{Gain of interferer transmit antenna}                & \multicolumn{1}{c|}{$R_{I}$}               & Range between interferer and victim radar                                 \\ \hline
				\multicolumn{1}{|c|}{$\theta$ and $\phi$}  & \multicolumn{1}{c|}{Azimuth and elevation angle}                        & \multicolumn{1}{c|}{$g_{k}$}               & Received signal power for signal reflected by $k^{\text{th}}$ IRS element \\ \hline
				\multicolumn{1}{|c|}{$N$}                  & \multicolumn{1}{c|}{Number of IRS elements}                             & \multicolumn{1}{c|}{$\tau_{k}$}            & Time delay between reference and $k^{th}$ IRS element                     \\ \hline
				\multicolumn{1}{|c|}{$G_{e}$}              & \multicolumn{1}{c|}{Radiation gain of each IRS element}                           & \multicolumn{1}{c|}{$\sigma_{e}$}          & RCS of each IRS element                                                   \\ \hline
				\multicolumn{1}{|c|}{$\delta_{k}$}         & \multicolumn{1}{c|}{Phase shift introduced by each IRS element}         & \multicolumn{1}{c|}{$\sigma_{O}^{P}$}      & Overall RCS of passive IRS                                                \\ \hline
				\multicolumn{1}{|c|}{$\delta_{k}^{*}$}     & \multicolumn{1}{c|}{Optimal phase shift introduced by each IRS element} & \multicolumn{1}{c|}{$\sigma_{O}^{P^{*}}$}  & Optimized RCS of passive IRS                                              \\ \hline
				\multicolumn{1}{|c|}{$\sigma_{O}^{A^{*}}$} & \multicolumn{1}{c|}{Optimized PCS of active IRS}                        & \multicolumn{1}{c|}{$\Gamma$}              & Reflection gain of each active IRS elements                               \\ \hline
				\multicolumn{1}{|c|}{$G_{P}$}              & \multicolumn{1}{c|}{Processing gain}                                    & \multicolumn{1}{c|}{$\rho$}                & Noise power at receiver                                                   \\ \hline
				\multicolumn{1}{|c|}{$k$}                  & \multicolumn{1}{c|}{Boltzmann's constant}                               & \multicolumn{1}{c|}{$F_{n}$}               & Noise Figure                                                              \\ \hline
				\multicolumn{1}{|c|}{$T_{0}$}              & \multicolumn{1}{c|}{receiver temperature}                               & \multicolumn{1}{c|}{$P_{D}$}               & Probability of detection for CA-CFARD algorithm                           \\ \hline
		\end{tabular}}
		\label{tab:tabel1p}
		\vspace{-5mm}
	\end{table}
	\subsection{Received FMCW Signal Model}
	For a target vehicle located at range $R$ and with relative velocity $\nu$, the received  FMCW frame is given as \cite{rad_dm}
	\begin{equation}
		\resizebox{0.9\columnwidth}{!}{$x_{R}\left(t\right) = \sqrt{P_{r}} \sum\limits_{l = 0}^{N_{f}-1}s_{b}\left(t-\tau_{l}-lT_{r}\right) \cdot \text{rect}\left(\frac{t-\tau_{l}-T{r}/2 -lT_{r}}{T_{r}}\right)+ n(t)$}
	\end{equation}
	where $\tau_{l} = \frac{2\left(R+\nu t+\nu lT_{r}\right)}{c_{0}}$ represents round-trip-time of the $l^{\text{th}}$ chirp signal in the FMCW frame from the target vehicle, $n\left(t\right)$ denotes the additive white Gaussian noise (AWGN) and $P_{r}$ represents the received signal power given as \cite{rad_adv}
	\begin{equation}
		\resizebox{0.31\columnwidth}{0.04\columnwidth}{$P_{r} = \frac{P_{t}G_{t}G_{r}\sigma\lambda^{2}}{\left(4\pi\right)^{3}R^{4}}$}
		\label{eq:ptl}
	\end{equation}
	here $G_{t}$, and $G_{r}$ respectively denote the gain of transmitting and receiving radar antennas and $\sigma$ RCS of the target. The received FMCW frame $x_{R}\left(t\right)$ is de-chirped by the following steps i.) Multiplying each received chirp signal with the reference chirp signal $x_{b}\left(t\right)$ and ii.) Low pass filter (LPF) with cut-off frequency $B_{c}$. The de-chirped FMCW frame is expressed as \cite{rad_dm}
	\begin{equation}
		\resizebox{0.89\columnwidth}{!}{$x_{V}\left(t\right) = \sqrt{P_{r}}\exp{\left(j4\pi f_{c}/c_{0}\right)}\sum\limits_{l = 0}^{N_{f}-1}\exp{\left\lbrace j2\pi \left(f_{b}t+f_{D}lT_{r}\right)\right\rbrace} + n_{V}\left(t\right)$}
		\label{eq:rdmap}
	\end{equation}
	where $f_{b} = \frac{2S_V R}{c_{0}}$ and $f_{D} = \frac{2f_{c}\nu}{c_{0}}$ represent beat and Doppler frequency respectively, and $n_{V}\left(t\right)$ is the noise component at the output of the LPF. 
	The range $R$ and relative velocity $\nu$ of the target are determined by the following steps i.) Computing the two dimensional Fast Fourier transform of \eqref{eq:rdmap}, which is called as the range-doppler (R-D) plot, and ii.) Applying thresholding algorithms such as variants of constant false alarm rate (CFAR) on the R-D to determine the target parameters. It is important note here the CFAR algorithms assume the knowledge of noise statistics \cite{rad_book}.
	\section{Interference in Automotive radar}
	\begin{figure}
		\centering
		\includegraphics[height = 2.6cm,width=0.6\columnwidth]{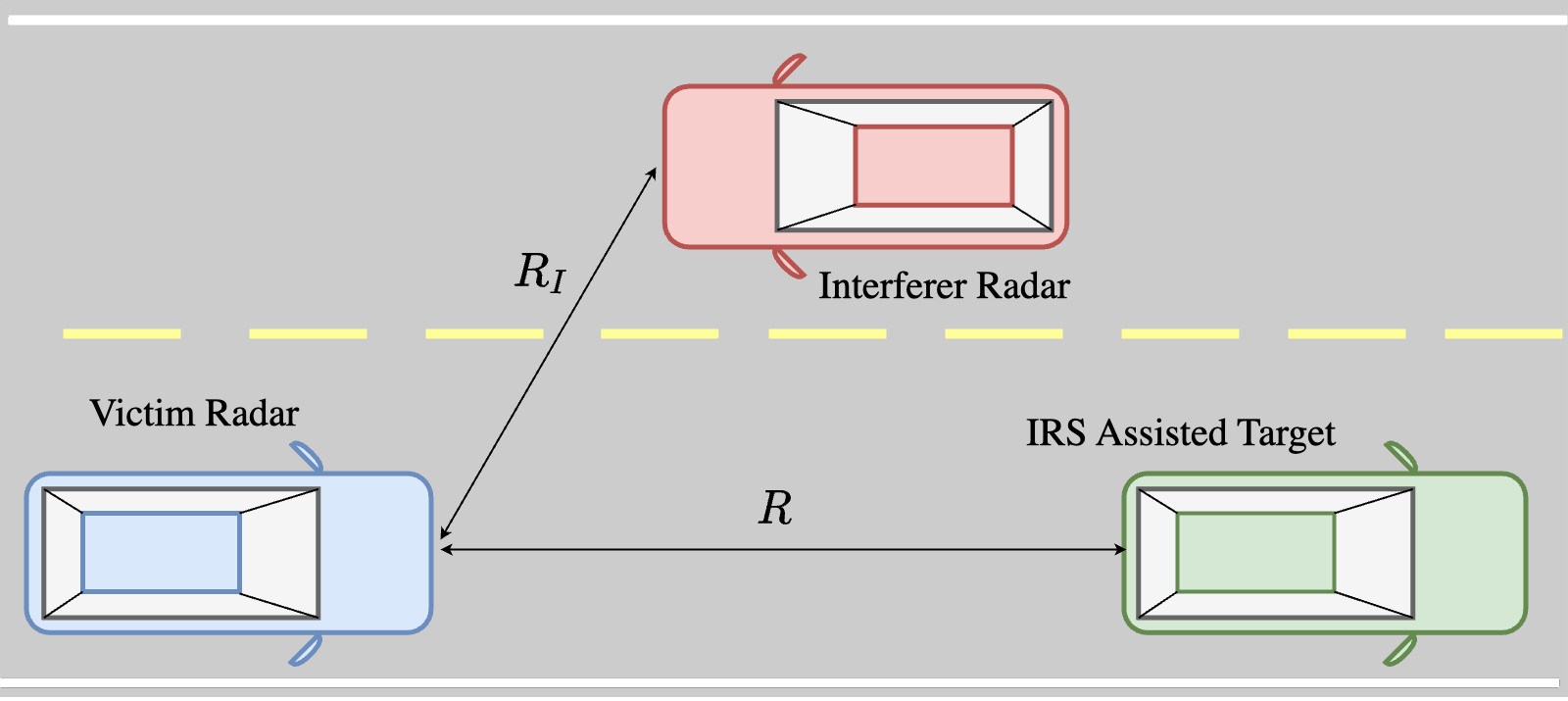}
		\caption{Radar interference scenario with IRS assisted target.}
		\label{fig:scenario}
		\vspace{-3mm}
	\end{figure}
\begin{figure}
		\begin{subfigure}{0.48\columnwidth}
			\includegraphics[height = 2.3cm,width=\columnwidth]{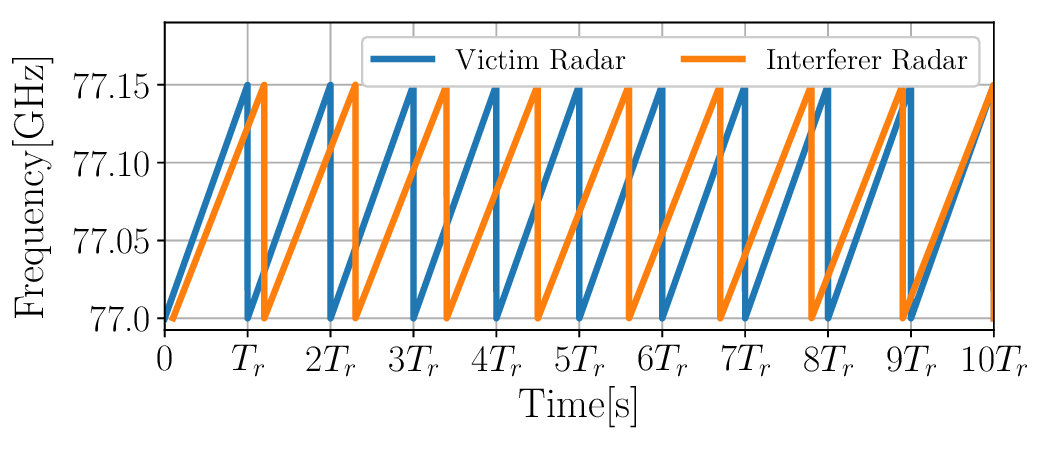}
			\caption{Similar slope interference.}
			\label{fig:similarslope1}
		\end{subfigure}
		\begin{subfigure}{0.48\columnwidth}
			\includegraphics[height = 2.3cm,width=\columnwidth]{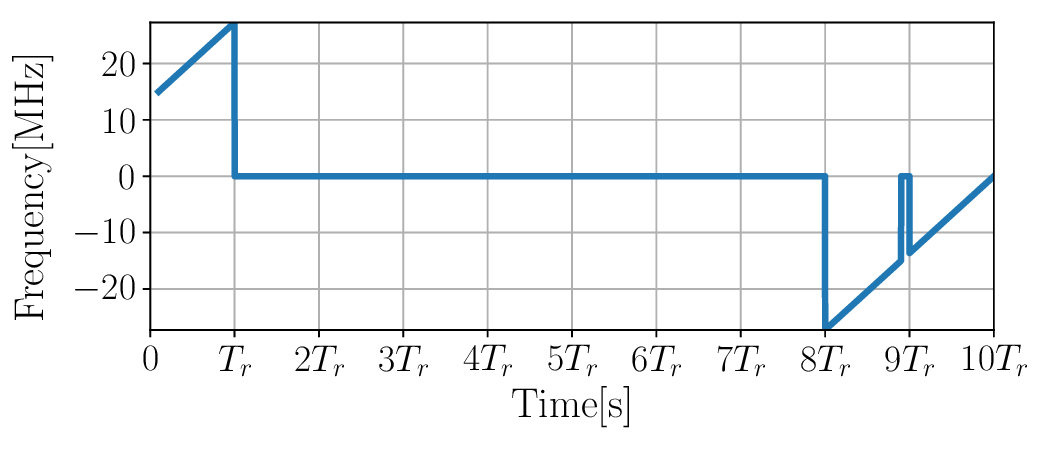}
			\caption{LPF output at victim radar.}
			\label{fig:similarslope2}
		\end{subfigure}
		\caption{Similar slope interference with $S_{V} = 20.45~\text{MHz}/\mu s$  and $S_{V} = 1.1S_{I}$}
		\label{fig:similarslope}
		\vspace{-4mm}
	\end{figure}
\begin{figure}
		\begin{subfigure}{0.48\columnwidth}
			\includegraphics[height = 2.3cm,width=\columnwidth]{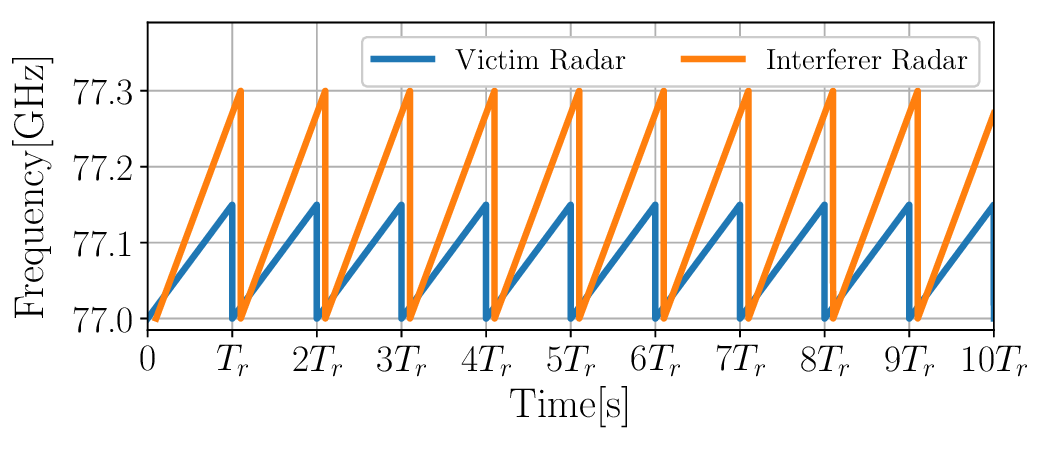}
			\caption{Sweeping slope interference.}
		\end{subfigure}
		\begin{subfigure}{0.48\columnwidth}
			\includegraphics[height = 2.3cm,width=\columnwidth]{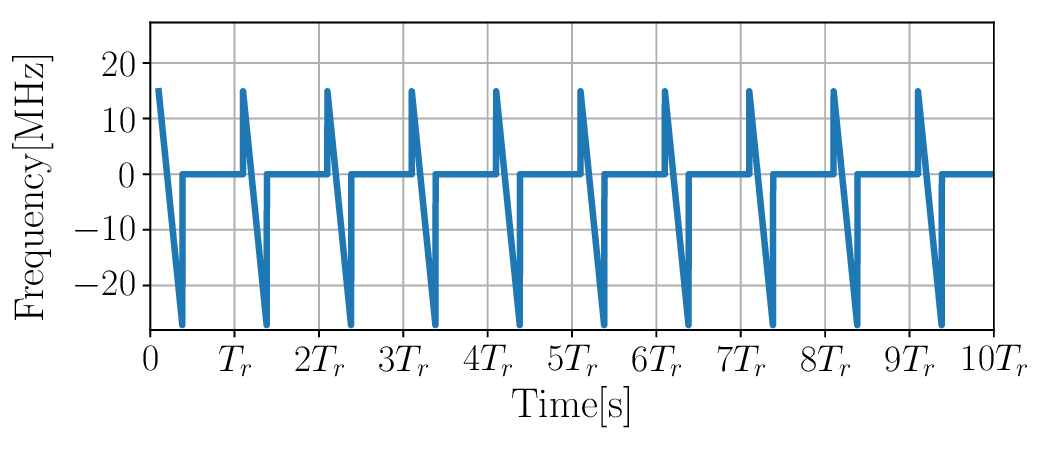}
			\caption{LPF output at victim radar.}
		\end{subfigure}
		\caption{Sweeping slope interference with $S_{V} = 20.45~\text{MHz}/\mu s$ and $S_{I} = 40.90~\text{MHz}/\mu s$.}
		\label{fig:sweepingslope}
	\end{figure}

	Fig.\ \ref{fig:scenario} illustrates the radar interference scenario wherein the victim radar measures the target vehicle parameters in the presence of interferer radar located at range $R_{I}$. With lack of standardization and co-ordination, parameters of FMCW based automotive radars vary with OEMs. Hence the interference between automotive radars from different or same OEMs are categorized as \cite{diff_int}:
	
	i.) \textit{Similar slope interference:} This type of interference occurs when both the victim and the interfering radars choose to operate with the same bandwidth and chirp duration leading to identical chirp slopes. However, due to hardware `imperfection', they cannot produce identical slopes in practice. Rather, the slopes are expected to be slightly different and vary within $10\%$, hence is referred to as the \textit{similar slope} interference. Fig.\ \ref{fig:similarslope} illustrates the similar slope interference scenario with victim and interfering radar operating at slopes $S_{V}$ and $S_{I}$, respectively. Chirp duration of the victim radar is denoted as $T_{r}$. From the output of LPF at the victim radar in Fig.\ \ref{fig:similarslope2}, it is observed that the interference is dominant across the entire chirp duration in certain chirps within the FMCW frame.
	
	ii.) \textit{Sweeping interference:} This type of interference arises when the victim and the interfering radars are operating with significantly different slopes because they have chosen to operate with different bandwidth and (or) chirp duration. Interference scenario for the victim and the interfering radars operating at different slopes is shown in Fig.\ \ref{fig:sweepingslope}. Here it is observed that interference is present across all chirps but affecting only for a small fraction of the chirp duration. 
	
	In this work, we consider both types of interference when evaluating the performance of the proposed IRS-assisted automotive radar. 
	\vspace{-2mm}
	\section{Automotive Radar Interference Model}
	Consider an interferer radar chirp signal $x_{I}\left(t\right)$ received at a time offset $\tau_{I}$ with respect to the reference chirp signal of victim radar. The output of LPF $x_{O_{l}}\left(t\right)$ for the $l^{\text{th}}$ received chirp signal $x_{V_{l}}\left(t\right)$ from the target is given as \cite{rad_lit4}
	\begin{equation}
		x_{O_{l}}\left(t\right) = \left\{\begin{matrix}
			x_{V_{l}}\left(t\right) + x_{I}\left(t\right);&  t \in T_{u} \\x_{V_{l}}\left(t\right); &  \text{Otherwise}\\
		\end{matrix}\right.  
	\end{equation}
	here $T_{u}$ is the vulnerable time period in which the frequency content $\mathcal{F}\left \lbrace x_{V_{l}}\left(t\right) + x_{I}\left(t\right) \right\rbrace$ lies within the bandwidth of the LPF. Further, $x_{I}\left(t\right)$ is given as  
	\begin{equation}
		\resizebox{0.7\columnwidth}{0.04\columnwidth}{$x_{I}\left(t\right) = \sqrt{P_{r_{I}}} \exp\left\lbrace j2\pi\left[\frac{1}{2}\left(S_{V}t^{2} - S_{I}\left(t-\tau_{I}\right)^{2}\right)\right] \right\rbrace$}
	\end{equation}
	here $S_{I}$ is the slope of the interferer radar and $P_{r_{I}}$ represents the received interference power and is of the form \cite{rad_adv}
		\begin{equation}
			\resizebox{0.325\columnwidth}{0.05\columnwidth}{$P_{r_{I}} = \frac{P_{t_{I}} G_{t_{I}} G_{r} \lambda^{2}}{\left(4\pi R_{I}\right)^{2}}$}
			\label{eqn:pti}
		\end{equation}
		here $P_{t_{I}}$ and $G_{t_{I}}$ respectively denote the transmitter power and gain of the interferer radar, and $R_{I}$ is the range between the interferer and the victim radars. Now comparing \eqref{eq:ptl} and \eqref{eqn:pti}, it is observed that interference power $P_{r_{I}}$ will be significant as compared to received power $P_{r}$ from the target for small interferer range $R_{I}$ and large target range $R$.
		\begin{figure}
			\centering
			\includegraphics[width=\columnwidth]{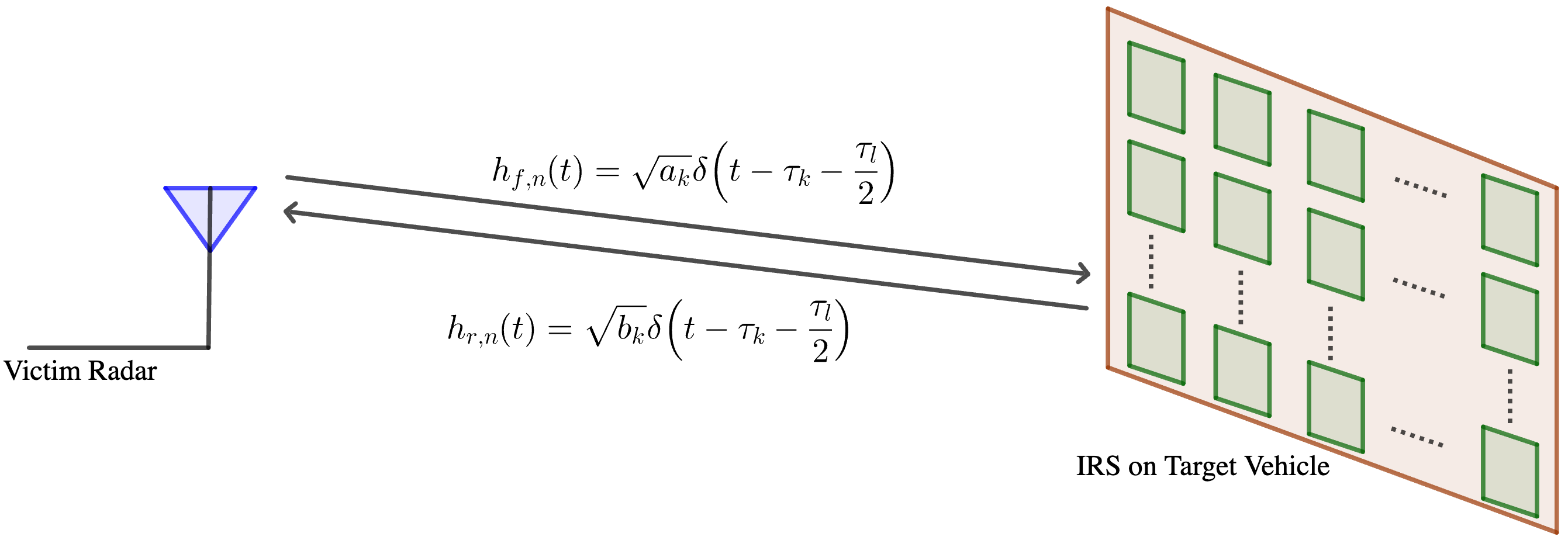}
			\caption{Channel response between victim radar and IRS assisted target.}
			\label{fig:detail}
		\end{figure}
		\section{IRS Assisted Enhanced Target Detection}
		Consider a passive IRS with $N$ elements embedded on the rear side of the target vehicle as shown in Fig.\ \ref{fig:detail}. Further, the victim radar is assumed to be at azimuth angle $\theta$ and elevation angle $\phi$ from the IRS-assisted target vehicle. For a passive IRS with $N$ elements, the received signal at the victim radar is
			\begin{equation}
			x_{V}^{\text{P-IRS}}\left(t\right)  = \sum\limits_{k=0}^{N-1} \left\lbrace h_{f,k}\left(t\right) * x_{T}\left(t\right) * h_{r,k}\left(t\right) \right \rbrace \exp(j \delta_{k})
			\label{eqn:oreq}
		\end{equation}
		 here $\delta_{k}$ denotes the phase shift introduced by each IRS element, the channel impulse response along forward and reflected path is represented as $h_{f,k}\left(t\right) $ and $h_{r,k}\left(t\right)$ respectively and is given as
		 \begin{equation}
		 	h_{f,k}\left(t\right) = \sqrt{a_{k}} \delta\left(t-\tau_{k}-\frac{\tau_{l}}{2} \right) 
		 	\label{eqn:fwdp}
		 \end{equation} 
		 		 \begin{equation}
		 	h_{r,k}\left(t\right) = \sqrt{b_{k}} \delta\left(t-\tau_{k}-\frac{\tau_{l}}{2} \right)
		 	\label{eqn:revp} 
		 \end{equation} 
		 here the channel gain along forward path is $a_{k} = \sqrt{\frac{G_{t}}{4\pi R^{2}}\sigma_{e}}$,  with $\sigma_{e} = \frac{\lambda^{2}G_{e}^{2}}{4\pi}$ representing the RCS  of each IRS element and reflection gain of each IRS element $G_{e} = \pi$ \cite{simris}. The time delay $\tau_{k}$ between the $k^{\text{th}}$ and the reference IRS element with position co-ordinates $\left(x_{k},y_{k} \right) $ is given as
		 \begin{equation}
		 	\tau_{k} =  \frac{1}{c_{0}}\left[x_{k}\sin\left(\theta\right)\cos\left(\phi\right) +y_{k}\sin\left(\theta\right)\sin\left(\phi\right)\right]
		 \end{equation}
		 Substituting \eqref{eqn:fwdp} and \eqref{eqn:revp} in \eqref{eqn:oreq}, the received signal at victim radar can be written as
		 \begin{equation*}
		 	x_{V}^{\text{P-IRS}}\left(t\right)  = \sum\limits_{l=0}^{N_{f}-1}\sum\limits_{k=0}^{N-1}\left\lbrace \sqrt{a_{k}b_{k}} s_{b}\left(t-2\tau_{k}-\tau_{l}-lT_{r}\right)\right.
		 \end{equation*}
		 \begin{equation}
		 	\times \left.\exp{\left( j\delta_{k}\right) }\right\rbrace + n\left( t\right)
		 	\label{eqn:sbsig}  
		 \end{equation}
		 After de-chirping \eqref{eqn:sbsig}, i.e., the output of LPF $x_{V_{L}}^{\text{P-IRS}}\left(t\right) $ at the victim radar can be written as  
		 	\begin{equation*}
		 	x_{V_{L}}^{\text{P-IRS}}\left(t\right)  = \sum\limits_{l=0}^{N_{f}}\sum\limits_{k=0}^{N-1}\sqrt{a_{k}b_{k}} \exp\left\lbrace 2f_{c}\tau_{k} + f_{c}\tau_{l}-\frac{1}{2}S\left(2 \tau_{k}\right) ^{2}\right.
		 	\end{equation*}
		 	\begin{equation}
		 		\left. -\frac{1}{2}S\tau_{l}^{2} +2 S t \tau_{k}+St\tau_{l}-2S\tau_{k}\tau_{l}\right\rbrace \times \exp{\left(j\delta_{k} \right) }+ n_{V}\left(t\right)
		 		\label{eqn:lp1feq}
		 	\end{equation}
		 	Since $\tau_{k}^{2}\ll S$ and $\tau_{l}^{2}\ll S$, the terms $-\frac{1}{2}S\tau_{k}^{2}$, $-\frac{1}{2}S\tau_{l}^{2}$ and $S\tau_{k}\tau_{l}$ can be neglected and \eqref{eqn:lp1feq} is expressed as
		 		\begin{equation*}
		 		x_{V_{L}}^{\text{P-IRS}}\left(t\right)  = \sum\limits_{l=0}^{N_{f}}\sum\limits_{k=0}^{N-1}\sqrt{a_{k}b_{k}} \exp\left\lbrace4j f_{c}\tau_{k} + j f_{c}\tau_{l}\right.
		 	\end{equation*}
		 	\begin{equation}
		 		\left. +S t \left(1+\frac{2\tau_{k}}{\tau_{l}}\right)\tau_{l}\right\rbrace \times \exp{\left( j\delta_{k}\right) }+ n_{V}\left(t\right)
		 		\label{eqn:lpfeq}
		 	\end{equation}
		 	As $\tau_{l}\gg \tau_{k}$, \eqref{eqn:lpfeq} can be written as	
		 		 		\begin{equation*}
		 	x_{V_{L}}^{\text{P-IRS}}\left(t\right)  = \sum\limits_{l=0}^{N_{f}}\sum\limits_{k=0}^{N-1}\sqrt{a_{k}b_{k}} \exp\left\lbrace2 f_{c}\tau_{k} + f_{c}\tau_{l}\right.
		 \end{equation*}
		 \begin{equation}
		 	\left. +S t \tau_{l}\right\rbrace \times \exp{\left( j\delta_{k}\right) }+ n_{V}\left(t\right)
		 	\label{eqn:lp2feq}
		 \end{equation}
		 Further \eqref{eqn:lp2feq} can be simplified as
		 \begin{equation*} 
			\resizebox{\columnwidth}{!}{$x_{V}^{\text{P-IRS}}\left(t\right) = \exp\left\lbrace \frac{j4\pi f_{c}R}{c_{0}}\right\rbrace\sum\limits_{l=0}^{N_{f}-1}\sum\limits_{k=0}^{N-1} \left\lbrace \sqrt{a_{k}b_{k}} \exp{\left(j4\pi f_{c}\tau_{k}\right)}  \right.$}
		\end{equation*}
		\begin{equation}
			\exp{\left(j\delta_{k}\right)} \exp{\left\lbrace j2\pi \left(f_{b}t+f_{D}lT_{r}\right)\right\rbrace}+ n_{V}\left(t\right)
		\label{eqn:fineq}
		\end{equation} \\
		Further, due to the very small separation between the IRS elements, we assume $a_{k}b_{k} = g = \frac{P_{t}G_{t}G_{r}G_{e}^{2}\lambda^{4}}{\left(4\pi R\right)^{4}}$ for all $k$. Using \eqref{eqn:fineq}, the overall RCS$^{1}$ of the entire passive IRS is obtained as:
		\begin{equation}
			\sigma_{O}^{P} = \sigma_{e}\sum\limits_{k=0}^{N-1} \exp{\left(j4\pi f_{c}\tau_{k}\right)}\exp{\left(j\delta_{k}\right)} 
			\label{eqn:sig_ov_sim}
		\end{equation}
		$\bullet~$\textit{Phase optimization of IRS elements:}
		A fundamental goal of using IRS is to coordinate the phase shifts of each individual elements in a way so that the reflections of each element are combined coherently at the receiver for maximum gain. From \eqref{eqn:sig_ov_sim}, we can verify$^{1}$ that, the optimal phase shift at the $k^{\text{th}}$ IRS element is given as: 
		\begin{equation}
			\delta_{k}^{*} = -j4\pi f_{c} \tau_{k}
			\label{eqn:opt_phs}
		\end{equation}
		Thus, the optimal phase shift depends on the direction of arrival (DOA) angles $\theta$ and $\phi$ of the victim radar which can be estimated by various methods such as by using a planar antenna array at the target vehicle or allowing the victim vehicle to transmit its location to the target vehicle using vehicle-to-vehicle communication.
		Hence, by appropriate adjustment of the IRS element phases at the target vehicle, the received signal power at the victim radar can be improved by a factor of $N$ and the optimised RCS of a passive IRS-assisted target vehicle is computed as: 
		\begin{equation}
			\sigma_{O}^{P^{*}} = N\sigma_{e}
			\label{eqn:eff_rcs}
			\vspace{-1mm}
		\end{equation}
		$\bullet~$\textit{RCS of active IRS:}
		It will be shown later in the results section, that passive IRS is not a practical option for improving the effective RCS of the target vehicle. However, by considering active IRS, i.e., by providing a \textit{reflection gain}, $\Gamma$, at each IRS element, the effective RCS can be increased to: 
		\begin{equation}
			\sigma_{O}^{A^{*}} = \Gamma N\sigma_{e}
			\label{eqn:acris}
		\end{equation}
		The output of LPF due to the signal reflected by active IRS-assisted target vehicle is obtained as:  
		\begin{equation}
			\resizebox{\columnwidth}{!}{$x_{V}^{\text{A-IRS}}\left(t\right) = \exp\left\lbrace j4\pi f_{c}R/c_{0}\right\rbrace\sum\limits_{l=0}^{N_{f}-1}\sum\limits_{k=0}^{N-1} \left\lbrace \Gamma~\sqrt{g_{k}} \exp{\left(j2\pi f_{c}\tau_{k}\right)} \cdot \right.$}
			\label{eqn:irs_opti_A}
		\end{equation}
		\begin{equation*}
			\left. \exp{\left(j\delta_{k}\right)} \exp{\left[ j2\pi \left(f_{b}t+f_{D}lT_{r}\right)\right]}+ n_{A_{k}} \left( t \right)\right\rbrace +n_{V}\left(t\right)
		\end{equation*}
		where the component $n_{A_{k}} \left( t \right)$ arises due to noise introduced by each active IRS element to its reflected signal \cite{ activeris}. The optimal phase shift $\delta_{k}$ at each element is determined according to \eqref{eqn:opt_phs}. 
		
		$\bullet~$\textit{Signal-to-interference ratio (SIR):}
		The ultimate goal of using the IRS in the target vehicle is to improve the SIR at the victim radar, which in turn can help improve the detection probability. 
		The SIR due to active IRS-assisted target is given as \cite{rad_dm}:
		\begin{equation}
			\resizebox{0.5\columnwidth}{0.04\columnwidth}{$\text{SIR}~\left[\text{dB}\right] = 10\log_{10}\left(\frac{G_{P}P_{a}}{P_{r_{I}}}\right)$}
			\label{eqn:sir}
		\end{equation}
		where $P_{a} = N\Gamma g^{2}$ represents the received signal strength and $G_{P} = T_{r}B_{c}N_{f}$ denotes the processing gain \cite{rad_adv}.
		\section{Results}
		In this section, we first compare the effective RCS of a typical vehicle when fitted with a passive vs. active IRS. This exercise reveals the interesting insight that the use of \textit{passive} IRS would actually \textit{reduce} the vehicle RCS instead of improving it and in turn motivates the need for using \textit{active} IRS. We then simulate the typical automotive radar interfering scenarios to obtain the reflection gain ($\Gamma$) that would be required for the active IRS to enable reliable target detection under automotive radar interference. 
		\subsection{Effective RCS: Passive vs. Active IRS}
		In~\cite{etsi}, the RCS of typical sedan cars is experimentally determined as $\sigma = 10~\text{dBm}^{2}$, which will be used as the \textbf{baseline} RCS and referred to as the \textbf{non-IRS} scenario in this paper. It can be assumed that the useful reflections from the target vehicle actually comes from the metals at the rear of the vehicle covering an area of $1m \times 1m$, approximately. If this area is covered with an IRS, we could fit about $256\times 256 = 65,536$ IRS elements~\footnote{Assuming each element has a size of $\lambda/2~\text{m} \times \lambda/2~\text{m}$ and a spacing of $\lambda/2~\text{m}$ between the elements, where $\lambda = 3.89mm$ for 77GHz radars.}. Thus, the effective RCS for passive IRS can be computed as 0.78sm or -1.079 dBsm according to Eqn. (\ref{eqn:eff_rcs}). 
		
\begin{figure}
			\centering
			\includegraphics[width=0.5\columnwidth]{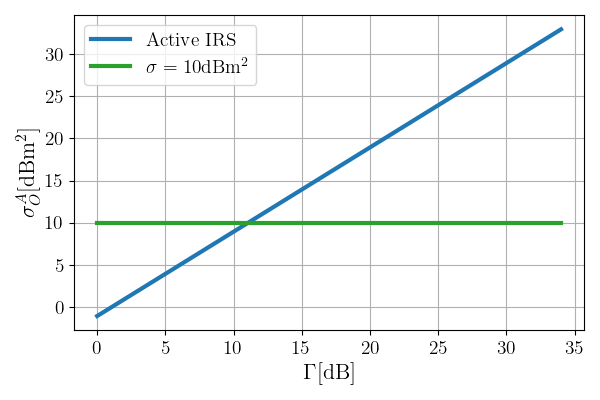}
			\caption{RCS improvement with active IRS as a function of its reflection gain, $\Gamma$.}
			\label{fig:sigma_eff_gamma_v2}
		\end{figure}

		This interesting result shows that \textit{simply covering the rear of the vehicle with a \textbf{passive} IRS will not be effective and would actually significantly reduce the RCS of the vehicle from its baseline value}. This result further motivates us to consider \textbf{active} IRS, which holds the promise to increase the RCS beyond the baseline by configuring adequate reflection gain ($\Gamma$) for the IRS elements. For an active IRS, the improvement in effective RCS as a function of $\Gamma$ is shown in Fig.\ \ref{fig:sigma_eff_gamma_v2}. We find that active IRS can overcome the baseline RCS with $\Gamma > 10~\text{dBm}^{2}$. 
		
		Improvement in RCS will result in improvement in SIR, which in turn will improve the detection probability. Therefore, from Fig.\ \ref{fig:sigma_eff_gamma_v2}, we can deduce that active IRS can potentially improve the detection probability compared to the non-IRS case if it is configured with $\Gamma > 10~\text{dBm}^{2}$.  However, we still do not know the required value of $\Gamma$ that will enable highly reliable detection with a detection probability close to 100\%. In the following, we conduct simulations to evaluate the probability of detection as a function of $\Gamma$ to find this answer for different types of FMCW interference, i.e., \textit{similar slope} vs. \textit{sweeping slope}.
		\subsection{Simulation Parameters and Performance Metrics}
		\begin{table}[]
			\caption{Simulation Parameters}
			\centering
			\begin{tabular}{|cc|}
				\hline
				\multicolumn{2}{|c|}{Victim FMCW Parameters} \\ \hline
				\multicolumn{1}{|c|}{Measurement Parameters} & FMCW Frame Parameters \\ \hline
				\multicolumn{1}{|c|}{$R_{max}$ = $200~\text{m}$} & $T_{r} = 5.5\times\frac{2R_{max}}{c_{0}} = 7.33~\mu\text{s}$ \\ \hline
				\multicolumn{1}{|c|}{\multirow{2}{*}{$\Delta R$ = $1~\text{m}$}} & $B_{s} = 150~\text{MHz}$ \\ \cline{2-2} 
				\multicolumn{1}{|c|}{} & $S_{V}= 20.45~\text{MHz}/\mu \text{s}$ \\ \hline
				\multicolumn{1}{|c|}{$v_{max}$ = $70~\text{m/s}$} & $N_{f} = 128$ \\ \hline
				\multicolumn{1}{|c|}{$\Delta v$ = $2.075~\text{m/s}$} & $B_{c} = 27.27~\text{MHz}$ \\ \hline
				\multicolumn{2}{|c|}{Transmitter and Receiver Parameters} \\ \hline
				\multicolumn{1}{|c|}{Transmit Power} & $1~\text{mW}$ \\ \hline
				\multicolumn{1}{|c|}{Antenna Gain} & $20~\text{dB}$ \\ \hline
				\multicolumn{1}{|c|}{Frequency} & $77~\text{GHz}$ \\ \hline
				\multicolumn{1}{|c|}{Noise Figure $F_{n}$} & $15~\text{dB}$ \\ \hline
				\multicolumn{1}{|c|}{System Temperature $T_{0}$} & $296^{\circ}K$ \\ \hline
				\multicolumn{1}{|c|}{CA-CFARD False Alarm Rate} & $10^{-5}$ \\ \hline
				\multicolumn{2}{|c|}{Target Parameters} \\ \hline
				\multicolumn{1}{|c|}{Range} & $180~\text{m}$ \\ \hline
				\multicolumn{1}{|c|}{Velocity} & $25~\text{m/s}$ \\ \hline
			\end{tabular}
			\label{tab:table2}
			\vspace{-5mm}
		\end{table}
		The parameters considered in the simulation are shown in Table \ref{tab:table2}. Cell averaging-CFAR detector (CA-CFARD) which is the basic version of CFAR is employed for detecting the target parameter from the R-D plot. The noise power at victim receiver is $\rho = kT_{0}B_{c}F_{n}$ where $k$ is Boltzmann's constant and noise power at each active IRS element is $\rho_{A} = kT_{0}B_{s}F_{n}$.
		
		The performance of target detection using active IRS is analysed in terms of probability of detection $P_{D}$, which is derived as the fraction of the total number of simulation trials that detected the target parameters correctly. The reflection gain of the active IRS, $\Gamma$, was increased in steps of $1~\text{dB}$ where 1000 trials were simulated for each $\Gamma$ by varying certain parameters randomly in each trial.  
		\begin{figure}[h]
			\centering
			\includegraphics[width=0.6\columnwidth]{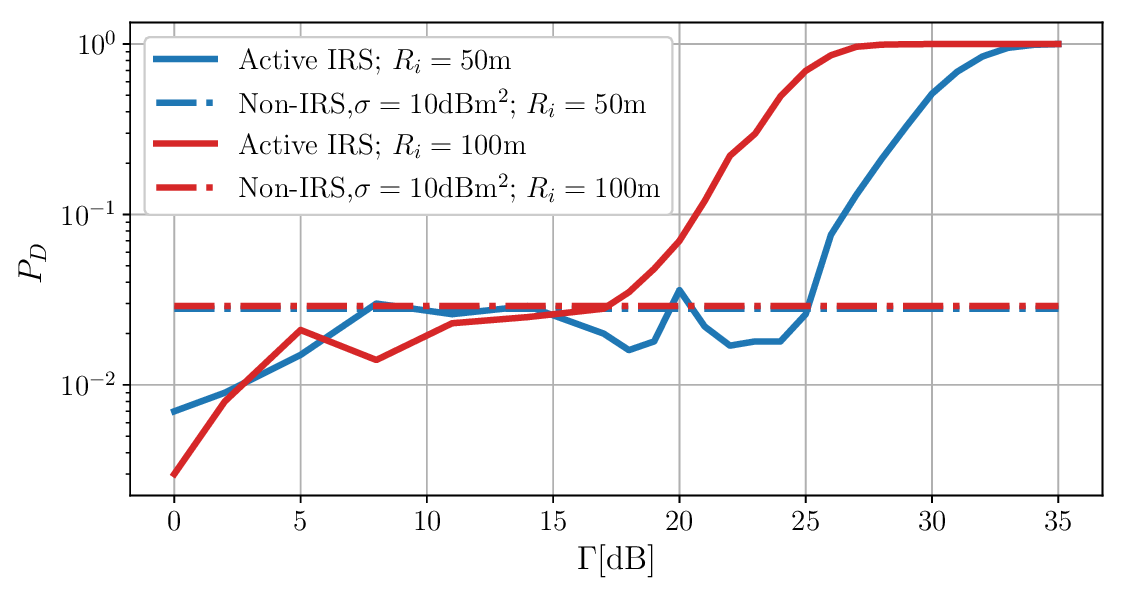}
			\caption{Detection probability for similar slope interference.
			}\label{fig:pd_gamma_similar_slope}
		\end{figure}
		\subsection{Detection Performance for Similar Slope Interference}
		\begin{figure*}[h]
			\centering
			\begin{subfigure}{0.16\textwidth}
				\includegraphics[width=\columnwidth]{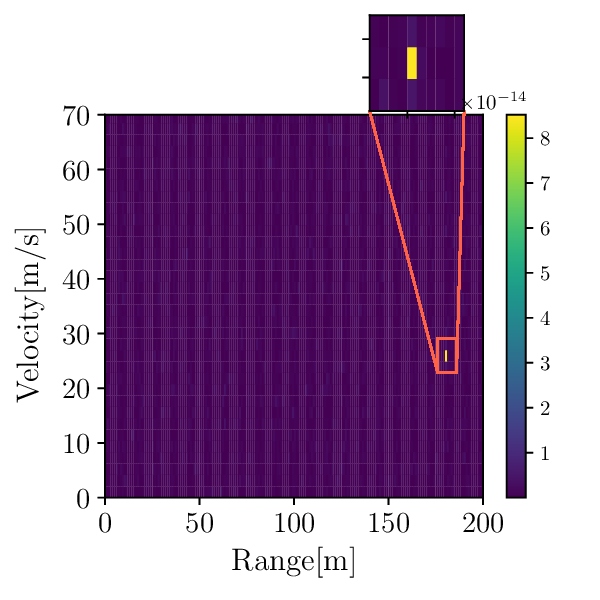}
				\caption{\scriptsize{R-D without\\ \hspace{\textwidth}interference and IRS.}}
			\end{subfigure}
			\begin{subfigure}{0.16\textwidth}
				\includegraphics[width=\columnwidth]{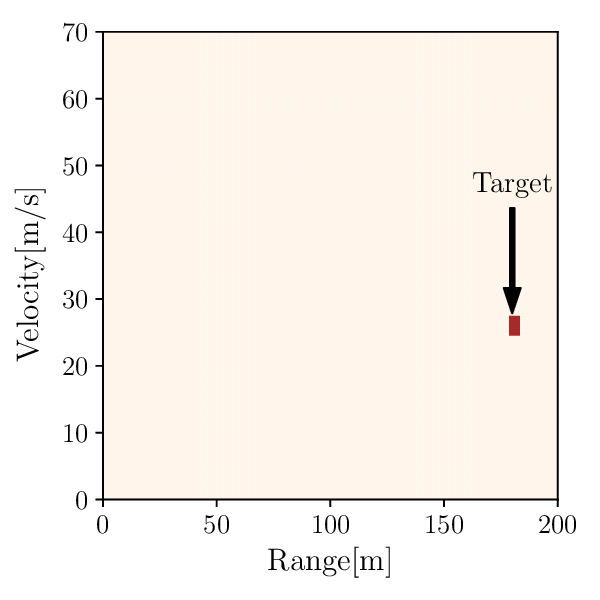}
				\caption{\scriptsize{CA-CFARD without\\ \hspace{\textwidth}interference and IRS.}}
			\end{subfigure}
			\begin{subfigure}{0.16\textwidth}
				\includegraphics[width=\columnwidth]{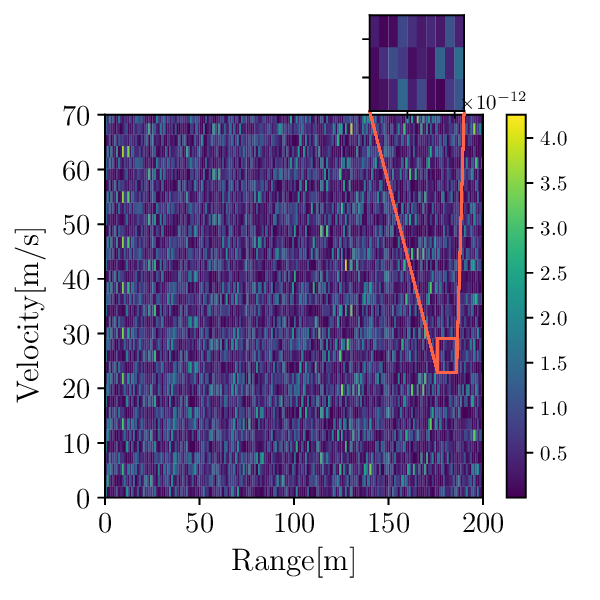}
				\caption{\scriptsize{R-D with $\Gamma=15~\text{dB}$.}}
			\end{subfigure}
			\begin{subfigure}{0.16\textwidth}
				\includegraphics[width=\columnwidth]{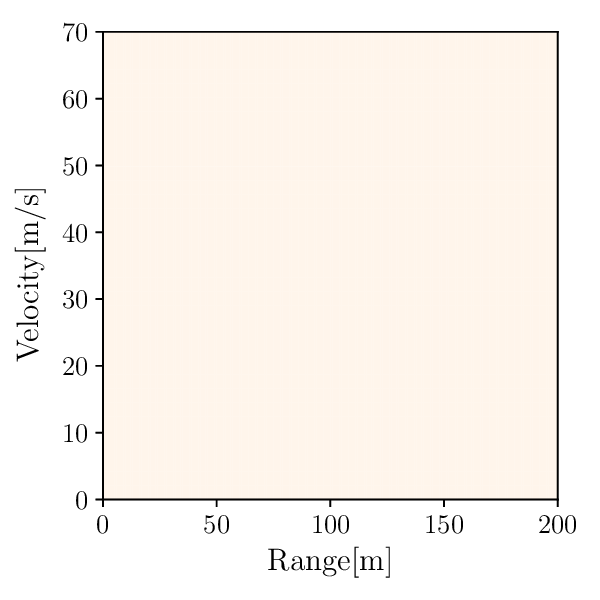}
				\caption{\scriptsize{CA-CFARD with \\ \hspace{\textwidth}$\Gamma=15~\text{dB}$.}}
			\end{subfigure}
			\begin{subfigure}{0.16\textwidth}
				\includegraphics[width=\columnwidth]{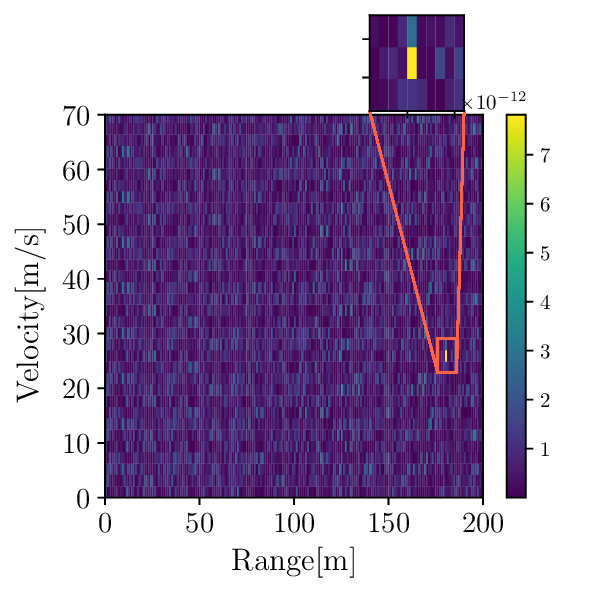}
				\caption{\scriptsize{R-D with $\Gamma=33~\text{dB}$.}}
			\end{subfigure}
			\begin{subfigure}{0.16\textwidth}
				\includegraphics[width=\columnwidth]{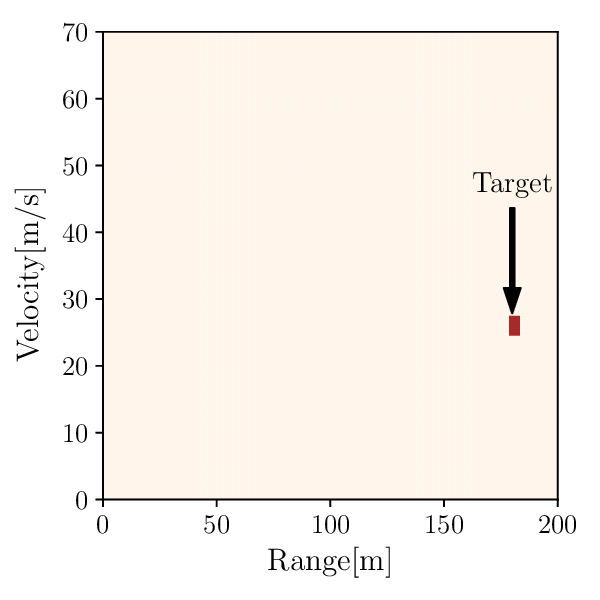}
				\caption{\scriptsize{CA-CFARD with\\ \hspace{\textwidth}$\Gamma = 33~\text{dB}$.}}
			\end{subfigure}
			\caption{Comparison of R-D plot and CA-CFARD for $\Gamma = 15~\text{dB}$ and $33~\text{dB}$. The range of interferer radar $R_{i} = 50~\text{m}$.}
			\label{fig:rdp_ini_rev}
			\vspace{-5mm}
		\end{figure*}
		Fig. \ref{fig:pd_gamma_similar_slope} plots the detection probability for an active IRS assisted target as a function of the reflection gain, $\Gamma$, when the interfering radar is located at range $R_{I} = 50~\text{m}$ (more aggressive interference) and $R_{I} = 100~\text{m}$ (less aggressive interference). In the simulation, the slope of the interfering radar is randomly varied within the $\pm10\%$ of the victim radar and the time offset, $\tau_{l}$, is randomly varied between $0$ and the chirp duration $T_{r}$. 
		
		We make the following observations: i) The detection probability increases with the increase in reflection gain as expected, ii) target detection performance is significantly better when the interfering radar is located at a longer distances from victim radar (less aggressive interference), which is again expected, iii) For $R_{I} = 50~\text{m}$ and $R_{I} = 100~\text{m}$, $P_{D}=100\%$ is obtained for gain values 35 dB and 33 dB respectively, iv) for $\Gamma < 10~\text{dB}$ the target detection performance deteriorates compared to the non-IRS assisted target, which is due to the effective RCS being smaller than the RCS of non-IRS assisted target (See Fig.\ref{fig:sigma_eff_gamma_v2}), v) the $P_{D}$ of the non-IRS target is very similar (the two lines coincides in Fig. \ref{fig:pd_gamma_similar_slope}) for $R_{i} = 50~\text{m}$ and $R_{i} = 100~\text{m}$~\footnote{A possible reason for this outcome is because the detection performance depends on the number of samples affected by the interference. In case of similar slope interference, we have found that the significant number of samples in the entire frame are affected by the interference.}. 
		
\begin{figure}[h]
			\centering
			\includegraphics[width=0.7\columnwidth]{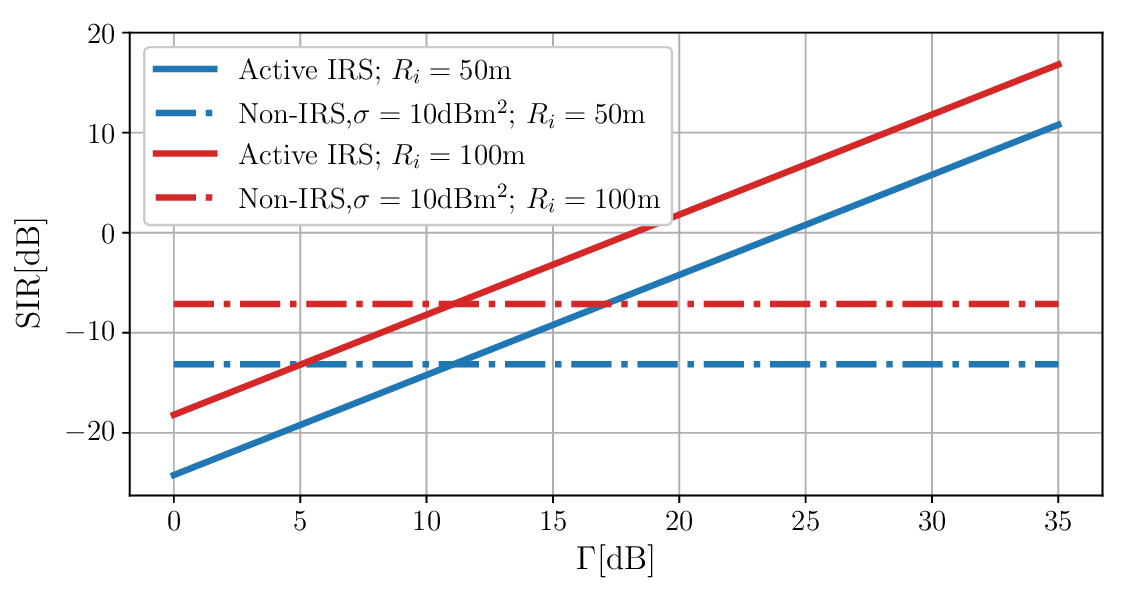}
			\caption{SIR as a function of reflection gain of the active IRS.
			}
			\label{fig:gamma_versus_sir_v2}
			\vspace{-3mm}
		\end{figure}

		The reason for improvement in detection performance with increasing $\Gamma$ is further explained using Fig. \ref{fig:rdp_ini_rev}, which shows the R-D plot and the corresponding CA-CFARD output for a single simulation trial. For $\Gamma = 33~\text{dB}$, we observe a clear peak and hence CA-CFARD is able to detect the target parameters. However for a smaller $\Gamma$, the interference will dominate and thereby CA-CFARD fails to detect the presence of the target. The reason for the peak (corresponding to target parameter) in the R-D plot becoming more prominent with increase in $\Gamma$ is investigated in Fig.\ \ref{fig:gamma_versus_sir_v2}, which shows that increase in $\Gamma$ significantly improves the SIR and this in turn results in a prominent peak at the target parameter location in the R-D plot. 
		\vspace{-2mm}
		\subsection{Detection Performance for Sweeping Slope Interference}
\begin{figure}[h]
			\centering
			\includegraphics[width=0.7\columnwidth]{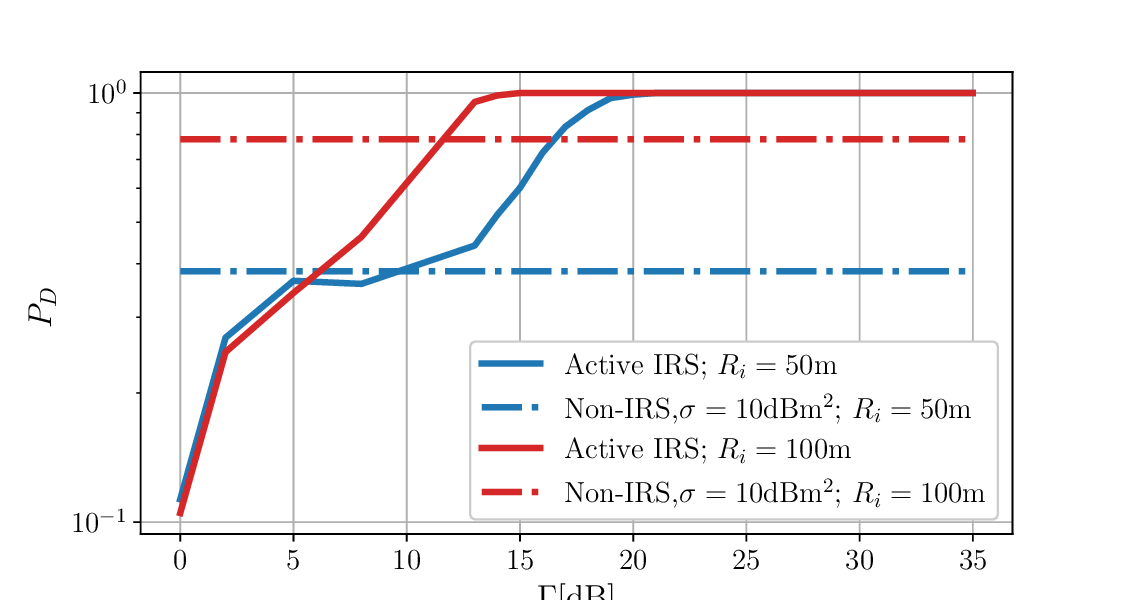}
			\caption{Detection probability for sweeping slope interference.
			}
			\label{fig:pd_gamma_sweeping_slope}
			\vspace{-5mm}
		\end{figure}

		Performance of IRS assisted target detection under sweeping slope interference with slope $40.90~\text{MHz}/\mu s$ is shown in Fig.\ \ref{fig:pd_gamma_sweeping_slope}, where the time offset $\tau_{I}$ is randomly varied between $0$ and the chirp duration $T_{r}$. Here it is observed that $100\%$ target detection is achieved for reflection gain of $10~\text{dB}$ earlier when to compared to similar slope interference. A possible reason for this outcome is due to only smaller fraction of victim chirp samples affected by interference (see Fig.\ \ref{fig:sweepingslope}). Further, for $R_{i} = 50~\text{m}$ and $R_{i} = 100~\text{m}$, $P_{D}=100\%$ is obtained for gain values 20 dB and 15 dB respectively.
		\vspace{-2mm}
		\section{Conclusion}
		\vspace{-1mm}
		This paper has proposed a novel solution for enhancing the performance of automotive radars under interference from other nearby radars. We have shown that an active onboard IRS can significantly increase a vehicle's RCS, thereby improving the strength of backscattered signals in the direction of the radar. Our simulations demonstrate that an active IRS with a reflection gain of 15-35 dB can achieve a probability of detection greater than 99\%, enabling even the most basic single-antenna radars to overcome various types of automotive radar interference. This paper also highlights the limitations of passive IRS, which could actually reduce the vehicle's effective RCS. Our findings have important implications for the design and development of automotive radar systems, and we believe that this proposed approach has the potential to revolutionize the field of vehicular sensing and communication. Future work will consider use of multi-antenna radars to further reduce the required reflection gain of the active IRS.
		\vspace{-2mm}
		

\begin{thebibliography}{00}
			\vspace{-2mm}
			\bibitem{rad_adv} S. Alland, W. Stark, M. Ali and M. Hegde, "Interference in Automotive Radar Systems: Characteristics, Mitigation Techniques, and Current and Future Research," in IEEE Signal Processing Magazine, vol. 36, no. 5, pp. 45-59, Sept. 2019.
			\bibitem{sae} SAE International, ``Surface Vehicle Information Report
			Active Safety System Sensors," J3088, Rev. Nov 2017.
			\bibitem{fmcw_pop} C. Waldschmidt, J. Hasch and W. Menzel, ``Automotive Radar — From First Efforts to Future Systems," in IEEE Journal of Microwaves, vol. 1, no. 1, Jan. 2021.
			\bibitem{mosarim} M. Kunert, ``The EU project MOSARIM: A general overview of project objectives and conducted work," 9th European Radar Conference, 2012. 
			\bibitem{simris}E. Basar and I. Yildirim, ``SimRIS Channel Simulator for Reconfigurable Intelligent Surface-Empowered Communication Systems," IEEE Latin-American Conference on Communications (LATINCOM), 2020. 
			\bibitem{activeris} Z. Zhang, L. Dai, X. Chen, C. Liu, F. Yang, R. Schober and H. Vincent Poor, ``Active RIS vs. Passive RIS: Which Will Prevail in 6G?," IEEE Transactions on Communications, vol. 71, no. 3, March 2023.
			\bibitem{radcom} C. Aydogdu, N. Garcia, L. Hammarstrand and H. Wymeersch, ``Radar Communications for Combating Mutual Interference of FMCW Radars," IEEE Radar Conference (RadarConf), 2019.
			\bibitem{rad_lit1} C. Fischer, H. L. Blöcher, J. Dickmann and W. Menzel, ``Robust detection and mitigation of mutual interference in automotive radar," 2015 16th International Radar Symposium (IRS), 2015.
			\bibitem{rad_lit2} J. Wang, ``CFAR-Based Interference Mitigation for FMCW Automotive Radar Systems," IEEE Transactions on Intelligent Transportation Systems, vol. 23, no. 8, Aug. 2022.
			\bibitem{rad_lit3} F. Jin and S. Cao, ``Automotive Radar Interference Mitigation Using Adaptive Noise Canceller," IEEE Transactions on Vehicular Technology, vol. 68, no. 4, April 2019.
			\bibitem{rad_lit4} F. Uysal and S. Sanka, ``Mitigation of automotive radar interference," IEEE Radar Conference (RadarConf18), 2018.
			\bibitem{rad_lit5} J. Rock, M. Toth, P. Meissner and F. Pernkopf, ``Deep Interference Mitigation and Denoising of Real-World FMCW Radar Signals," IEEE International Radar Conference (RADAR), 2020.
			\bibitem{rad_lit6} C. Jiang, Z. Zhou and B. Yang, ``Unsupervised Deep Interference Mitigation for Automotive Radar," 2022 IEEE 12th Sensor Array and Multichannel Signal Processing Workshop (SAM), 2022.
			\bibitem{rad_lit7} J. Rock, M. Toth, E. Messner, P. Meissner and F. Pernkopf, ``Complex Signal Denoising and Interference Mitigation for Automotive Radar Using Convolutional Neural Networks," 22th International Conference on Information Fusion, 2019.
			\bibitem{rs1} S. P. Maruthi, T. Panigrahi and M. Hassan, "Improving the Reliability of Pulse-Based Terahertz Communication using Intelligent Reflective Surface," 2020 IEEE International Conference on Communications Workshops (ICC Workshops), 2020.
			\bibitem{rs2}D. Ma, M. Ding and M. Hassan, "Enhancing Cellular Communications for UAVs via Intelligent Reflective Surface," 2020 IEEE Wireless Communications and Networking Conference (WCNC), 2020.
			\bibitem{rs2_2} Y. Chen et al., ``Downlink Performance Analysis of Intelligent Reflecting Surface-Enabled Networks," IEEE Transactions on Vehicular Technology (Early Access), 2022.
			\bibitem{rs3} A. M. Elbir, K. V. Mishra, M. R. B. Shankar and S. Chatzinotas, ``The Rise of Intelligent Reflecting Surfaces in Integrated Sensing and Communications Paradigms," arXiv preprint arXiv:2204.07265, April 2022.
			\bibitem{rs4} M. Rihan, E. Grossi, L. Venturino and S. Buzzi, ``Spatial Diversity in Radar Detection via Active Reconfigurable Intelligent Surfaces," arXiv preprint arXiv:2202.01616, May 2022.
			\bibitem{rs5} A. Shafie, G. N. Yang, C. Han, J. M. Jornet, M. Juntti and T. Kurner, ``Terahertz Communications for 6G and Beyond Wireless Networks: Challenges, Key Advancements, and Opportunities," arXiv preprint 	arXiv:2207.11021, July 2022.
			\bibitem{rad_lit8}S. K. Dehkordi and G. Caire, ``Making Vulnerable Road Users More Visible to Radar: A Communications Inspired Approach," 21st International Radar Symposium (IRS), 2021.
			\bibitem{rad_dm} V. Winkler, ``Range Doppler detection for automotive FMCW radars," 2007 European Microwave Conference, 2007.
			\bibitem{rad_book} B. M. Keel, ``Principles of Modern Radar: Basic principles", Chap. 16, IET Digital Library, 2010. 
			\bibitem{diff_int} R. Amar, M. Alaee-Kerahroodi and M. R. Bhavani Shankar, ``FMCW-FMCW Interference Analysis in mm-Wave Radars; An indoor case study and validation by measurements," 2021 21st International Radar Symposium (IRS), 2021.
			\bibitem{mathw} https://in.mathworks.com/help/driving/ref/vehicledimensions.html
			\bibitem{etsi} E. Bel Kamel, A. Peden and P. Pajusco, ``RCS modeling and measurements for automotive radar applications in the W band," 11th European Conference on Antennas and Propagation (EUCAP), 2017.
			\bibitem{hlt}H. L. Van Trees, Optimum Array Processing : Part IV, Detection, Estimation and Modulation Theory, John Wiley and Sons, 2002.
		\end{thebibliography}
	\end{document}